# No cold dust within the supernova remnant Cassiopeia A

Oliver Krause[1,2], Stephan M. Birkmann[2], George H. Rieke[1], Dietrich Lemke[2], Ulrich Klaas[2], Dean C. Hines[3] & Karl D. Gordon[1]

[1]*Steward Observatory, University Arizona, 933 N Cherry Ave, Tucson, AZ 85721, USA*

[2]*Max-Planck-Institut für Astronomie, Königstuhl 17, 69117 Heidelberg, Germany*

[3]*Space Science Institute, 4750 Walnut Street, Boulder, CO 80301, USA*

**A large amount (about three solar masses) of cold (18 K) dust in the prototypical type II supernova remnant Cassiopeia A was recently reported[1]. It was concluded that dust production in type II supernovae can explain how large quantities (~$10^8$ solar masses) of dust observed[2] in the most distant quasars could have been produced within only 700 million years after the Big Bang. Foreground clouds of interstellar material, however, complicate the interpretation of the submillimeter observations of Cas A. Here we report far-infrared and molecular line observations that demonstrate that most of the detected submillimetre emission originates from interstellar dust in a molecular cloud complex located in the line of sight between the Earth and Cas A, and is therefore not associated with the supernova remnant. The argument that type II supernovae produce copious amounts of dust is therefore not supported by the case of Cas A, which previously appeared to provide the best evidence for this possibility.**

It is well known that the Cas A supernova remnant is occulted by interstellar clouds of molecular gas[3], as was impressively demonstrated by the discovery of the OH molecule in absorption against it, the first radio detection of a molecule in interstellar space[4]. The visual extinction caused by dust associated with these clouds is probably the reason why only tentative historical records exist for the supernova outburst around AD 1680, although its proximity of 3.4 kpc would have made Cas A most likely the brightest stellar object in the sky[5]. The thermal emission of this foreground material should be easily detectable in submillimetre observations, given the high column densities of molecular gas ($N(H_2) > 10^{22}$ cm$^{-2}$) towards the supernova remnant[6]. Dunne et al. indeed note "an area of diffuse emission to the west of the remnant", which they however removed as being an instrumental artifact. Such extended emission was also noted independently by the observers who originally obtained the submillimetre data[7] using the Submillimetre Common-User Array (SCUBA)[8].

To clarify the nature of the submillimetre emission, we have observed Cas A at 160 μm using the Multiband Imaging Photometer (MIPS)[9] aboard the Spitzer Space Telescope. Not only does the spectral response of this band lie near the peak of the emission of cold interstellar dust, but the detectors are d.c.-coupled and operate in staring mode, so they respond well to extended emission. Our new measurements show an extended emission component (Fig. 1a and Fig. 2), distributed towards the southeast and west far beyond the forward shock that defines the outer boundary of the supernova remnant. The same large scale emission was independently detected at 170 μm using the ISOPHOT[10] Serendipity Survey[11] (Fig. 2).

In contrast, the chopping inherent in SCUBA data-taking suppresses extended emission; 1/$f$ noise in the bolometers can further complicate measurements of extended sources. Since different approaches to the data reduction can therefore yield different results about an extended background towards Cas A, we obtained the data used by Dunne et al.[1] from the SCUBA archive and re-reduced it. In our case, rather than forcing the baseline to zero, we determined it for each bolometer scan by subtracting the median level over the scan. The resulting submillimetre morphology after removal of the synchrotron emission (shown in Fig. 1a-c) is consistent with that observed at 160 μm.

Our submillimetre map confirms the presence of three knots of emission in the southeastern and western part of the remnant, which coincide with the emission detected by Dunne et al.[1] in their background subtracted map (compare Fig. 1d). We have not reanalyzed the SCUBA 450 µm data, but the image presented by Dunne et al. agrees well with their synchrotron-subtracted 850 µm image (Fig. 1d) and is dominated by the same knots. The 160 µm image also supports the presence of these knots, within the limitations of its lower angular resolution (beam diameter of 40" compared with 15" for the 850 µm map). From the ratio of the flux densities at 160 and 850 µm we derive a dust temperature of 14±2 K towards the dust clouds, a value commonly observed in dark clouds.

As shown in Fig. 1a, much of the 160 µm and 850 µm continuum emission coincides with the distribution of carbon-monoxide[12] emission from a molecular cloud complex at a radial velocity ($v_{LSR}$) of -50…-35 km/s with respect to the local standard of rest, which is kinematically associated with the Perseus spiral arm of our Galaxy. To extend the comparison of the dust emission with the shape of this molecular cloud to higher resolution, Figure 1c shows the optically thin OH absorption[13] by the molecular gas against the bright radio continuum background of Cas A. As the kinematics of the OH is similar to that of CO, it traces the same material. The correspondence of the OH absorption to the full resolution 850 µm emission image from Dunne et al.[1] (Fig. 1d) is extremely close, suggesting that much of the emission attributed by them to the Cas A supernova remnant instead arises in the foreground molecular material.

We can test quantitatively whether these molecular clouds could be responsible for the emission observed by Dunne et al.[1] by comparing the column densities of gas and dust towards the remnant. While a gas-to-dust mass ratio of about 100-200 is expected for the interstellar medium, pristine dust synthesized by the supernova is not expected to be associated with molecular gas, or at most in significantly smaller quantities. OH is preferred for this comparison because even $^{13}$CO is optically thick[3]. The total column density of gas was determined from the observed OH line opacity using an abundance[14,15] of $N(OH)/N(H_2) = 8 \times 10^{-8}$ and assuming that the excitation temperature equals the dust temperature of 14 K in these dense clouds. We derived the column density of dust from the optically thin 850 μm emission using a dust mass absorption coefficient[16] of $\kappa_{850 \mu m}=1.8$ cm$^2$ g$^{-1}$. Both quantities are highly correlated as shown in Fig. 3 (confidence level $P > 0.99$), corresponding to a gas-to-dust mass ratio of 120±24 in full agreement with the canonical value for the Milky Way. To bound the uncertainties, we used the synchrotron subtracted and baseline flattened 850 μm map from Dunne et al. to estimate a gas-to-dust ratio of 105±21. The good agreement with the value from our reduction of the same data shows that the result does not depend on the submillimeter data reduction procedure.

Assuming a distance of 3 kpc to the Perseus spiral arm clouds[3], the total masses of dust and gas projected across the face of the remnant are 14 $M_\odot$ and 1700$M_\odot$ respectively, where $M_\odot$ is the solar mass. Since the OH column density is seen in absorption against the radio emission of Cas A, the molecular material and dust accounting for most of the submillimetre emission must be located between the Earth and the remnant.

We can set an upper limit to the amount of cold dust in the supernova remnant using our 160 μm map, taken near the wavelength that should be most sensitive for its detection. After subtracting the interstellar foreground emission and the contribution from 0.003 $M_\odot$ of warm (82 K) dust[17] in Cas A, we derive an upper limit of 0.2 $M_\odot$ of cold dust. We have assumed protosilicate grains, which are suggested to account for the bulk of warm dust in the remnant. Since there is no evidence for cold (18 K) dust within Cas A, exotic explanations for the submillimetre signal are unnecessary. (Such explanations include the non-standard dust opacities as proposed by Dunne et al.[1] or Dwek's proposal for metallic needles[18].

The dust yield of Cas A, the prototypical Type II supernova, is therefore at least one order of magnitude lower than previously reported, weakening the argument that objects like it are the dominant source of interstellar dust. The presence of cold dust was recently also reported for the Kepler supernova remnant[19]. The generic type of this supernova is however controversial[20], with several lines of evidence pointing towards Type Ia[21,22]. In addition, the background/foreground emission toward the remnant is only slightly lower than for Cas A[23], raising the possibility that some of the submillimetre emission arises from material projected onto the remnant as in Cas A. Therefore, by itself the Kepler remnant does not make a compelling case that large amounts of dust are created in type II supernova remnants. Because of the dearth of other well-established type II supernova remnants, and the tendency for them to lie close to the Galactic plane where infrared and submm background/foreground emission is strong, it will be difficult to establish unequivocally the presence of large masses of cold dust in these objects.

**References**

1. Dunne, L., Eales, S., Ivison, R., Morgan, H. & Edmunds, M. Type II supernovae as a significant source of interstellar dust. *Nature* **424**, 285-287 (2003)

2. Bertoldi, F., Carilli, C.L, Cox, P. et al. Dust emission from the most distant quasars. *Astron. Astrophys.* **406**, L55-L58 (2003)

3. Wilson, T.L., Mauersberger, R., Muders, D. et al. The molecular gas toward Cassiopeia A. *Astron. Astrophys.* **280**, 221-230 (1993)

4. Weinreb, S., Barrett, A.H., Meeks, M.L. & Henry, J.C. Radio Observations of OH in the Interstellar Medium. *Nature* **200**, 829-831 (1963)

5. Hughes, D.W. Did Flamsteed See the Cassiopeia-A Supernova. *Nature* **285**, 132-133 (1980)

6. Reynoso, E.M. & Goss, W.M. Very Large Array Observations of 6 Centimeter $H_2CO$ in the Direction of Cassiopeia A. *Astrophys. J.* **575**, 871-885 (2002)

7. Loinard, L., Lequeux, J., Tilanus, R. & Lagache, P.O. Submillimeter Observations of Cassiopeia A. *Rev. Mex.A. A.* **15,** 267-269 (2003)

8. Holland W. S. *et al.* SCUBA: A common-user submillimetre camera operating on the James Clerk Maxwell Telescope. *Mon. Not. R. Astron. Soc.* **303**, 659–672 (1999)

9. Rieke, G.H., Young, E.T., Engelbracht, C. W. et al. The Multiband Imaging Photometer for Spitzer. *Astrophys. J. Supp.* **154,** 25-28 (2004)

10. Lemke, D., Klaas, U., Abolins, J. et al. ISOPHOT - capabilities and performance. *Astron. Astrophys.* **315**, L64-L70 (1996)

11. Bogun, S., Lemke, D., Klaas, U. et al. First data from the ISOPHOT FIR Serendipity survey. *Astron. Astrophys.* **315**, L71-L74 (1996)

12. Liszt, H. & Lucas, R. 86 and 140 GHz radiocontinuum maps of the Cassiopeia A SNR. *Astron. Astrophys.* **347**, 258-265 (1999)

13. Bieging, J.H. & Crutcher, R.M. VLA observations of 1667 MHz OH absorption toward Cassiopeia A. *Astrophys. J.* **310**, 853-871 (1986)

14. Crutcher, R.M. Nonthermal OH main lines and the abundance of OH in interstellar dust clouds. *Astrophys. J.* **234**, 881-890 (1979)

15. Harju, J., Winnberg, A. & Wouterloot, J.G.A. The distribution of OH in Taurus Molecular Cloud-1. *Astron. Astrophys.* **353**, 1065-1073 (2000)

16. Ossenkopf, V. & Henning, Th. Dust opacities for protostellar cores. *Astron. Astrophys.* **291**, 943 –959 (1994)

17. Hines, D.C., Rieke, G.H., Gordon, K.D. et al. Imaging of the Supernova Remnant Cassiopeia A with the Multiband Imaging Photometer for Spitzer (MIPS). *Astrophys. J. Supp.* **154,** 290-293 (2004)

18. Dwek, E. The Detection of Cold Dust in Cassiopeia A: Evidence for the Formation of Metallic Needles in the Ejecta. *Astrophys. J.* **607**, 848-854 (2004)

19. Morgan, H. L., Dunne, L., Eales, S. A., Ivison, R. J., & Edmunds, M. G. 2003, *Astrophys. J.*, **597**, L33-L36

20. Blair, W. P. in *1604-2004: Supernovae as Cosmological Lighthouses* (eds Turatto, M., Shea, W. R. J., Benetti, S. & Zampieri, L.) (ASP Conf. Ser., Astronomical Society of the Pacific, San Francisco) in the press) also available at (http://arxiv.org/abs/astro-ph/0410081)

21. Rothenflug, R., Magne, B., Chieze, J. P., & Ballet, J.: Hydrodynamic model of Kepler's supernova remnant constrained by EINSTEIN and EXOSAT X-ray spectra. *Ast. & Astrophys.*, **291**, 271-282 (1994)

22. Kinugasa, Kenzo; Tsunemi, Hiroshi: ASCA Observation of Kepler's Supernova Remnant. *Pub. Ast. Soc. Japan*, **51**, 239-252 (1999)

23. Arendt, R. G.: An infrared survey of Galactic supernova remnants. *Astrophys. J. Supp.*, **70**, 181-212 (1989)

24. Wright, M., Dickel, J., Koralesky, B. & Rudnick, L. The Supernova Remnant Cassiopeia A at Millimeter Wavelengths. *Astrophys. J*. **518**, 284-297 (1999)

**Correspondence** and request for materials should be addressed to O.K. (krause@as.arizona.edu)

**Acknowledgements** We thank Harvey Liszt and John Bieging for making their molecular line observations available for us. This work is based on observations made with the Spitzer Space Telescope and the Infrared Space Observatory ISO. We acknowledge access to the SCUBA data archive operated by CADC. We would like to thank Loretta Dunne, Haley Morgan, Steve Eales and Rob Ivison for helpful discussions.

**Competing interests statement** The authors declare that they have no competing financial interests.

Figure 1: Continuum and molecular line emission from dust and gas. (**a-c**) Nyquist-sampled SCUBA 850 μm images after removal of the synchrotron component by subtracting a 3.7 mm image[24] scaled to match the extrapolated[11] synchrotron flux of 35 Jy at 850 μm. The white circles indicate the position of the reverse and forward shocks estimated from X-ray observations. The total flux of integrated 850 μm dust emission within the forward shock is 16.0 ± 2.4 Jy. The contours overlaid are 160 μm surface brightness (black, increment 30 MJy/sr) and integrated intensity of $^{13}$CO *J*=1-0 emission (blue, increment 2.5 K km/s) (**a**), 850 μm surface brightness (increment 20 mJy/beam, starting 95 mJy/beam) (**b**) and integrated optical depth of OH $^2\Pi_{3/2}$ *F*=2-2 absorption (increment 0.1) (**c**). Beam sizes are indicated (circles). The OH and 3.7 mm observations have been convolved to match the 15" SCUBA beam and were sampled at the same positions. (**d**) The baseline-flattened and synchrotron subtracted 850 μm map from Dunne et al. with the same countour levels as **b**) for comparison. All panels show an area of 12'x12' centered at the remnant.

Figure 2: Large scale far-infrared emission from dust. The data have been obtained using MIPS and the ISOPHOT Serendipity Survey (ISOSS). The ISOPHOT 170 μm map in the background shows a 40'x80' large region centered on Cas A. Colours represent 170 μm surface brightness (range 140 - 300 MJy/sr). The MIPS 160 μm map (convolved to the 100" ISOSS beam) is overlaid as contours. An issue with the MIPS data is that the integration ramps saturate on the bright background around Cas A. However, the individual

amplifier reads at 8/second are sent to the ground, and we found it possible to extract meaningful measurements from the first few reads prior to saturation. The quality of the data was confirmed by comparing integrated fluxes with those from ISO (the agreement was within the calibrational uncertainties of ~20%) and through laboratory measurements that show that the recovery from saturation is quick without serious implications for the following integration. The larger field of the ISOSS map reveals the presence of dust emission extending considerably beyond the field shown in Fig. 1 (white box).

Figure 3: Gas-to-dust correlation. The column densities of dust and gas towards Cassiopeia A were determined point by point from the Nyquist-sampled 850 µm and OH maps within the outline shown in Fig. 1c, where the radio brightness was high enough to measure the strength of the OH absorption accurately. The outer boundary of this area coincides with the forward shock of the remnant. The error bars of the individual measurements (corresponding to the 1σ noise in the 850 µm and OH opacity maps) are indicated in the upper left. The data can be fitted to a straight line with a slope of 120±24. The quoted 1σ uncertainty of the gas-to-dust mass ratio is a combination of the fitting uncertainty of the slope (which accounts for noise in the maps and temperature uncertainties) and the calibration errors (~10% at 850 µm and ~15% for the OH measurements).

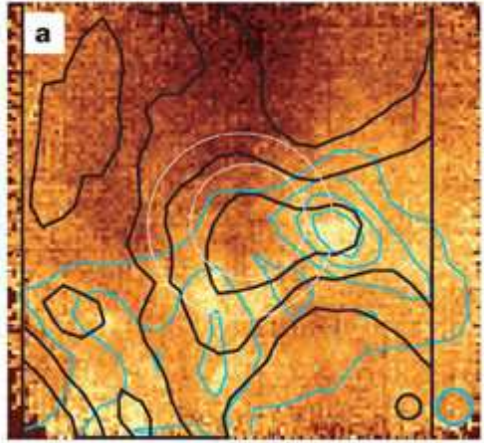
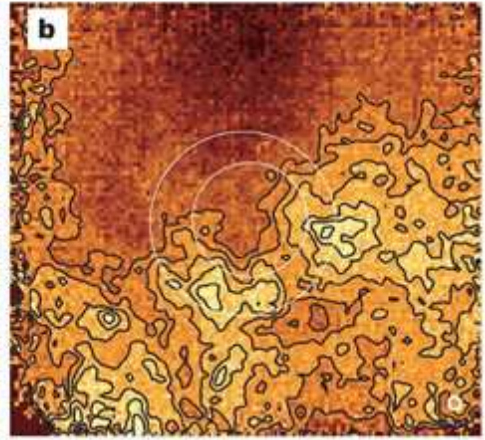
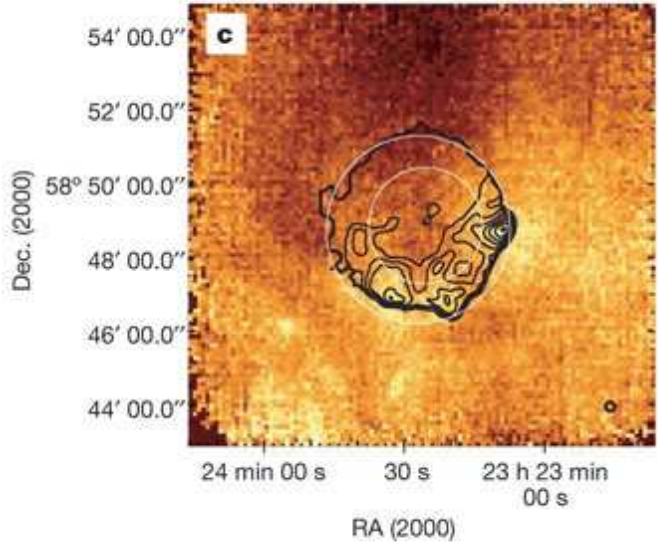
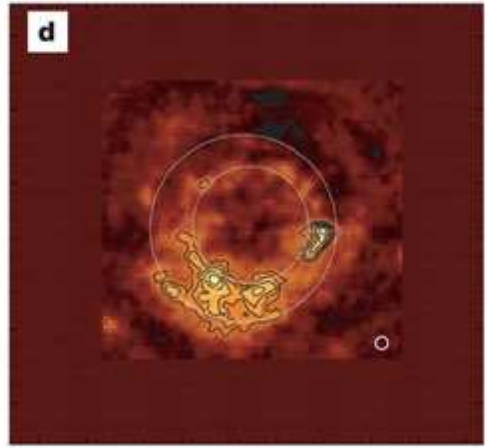

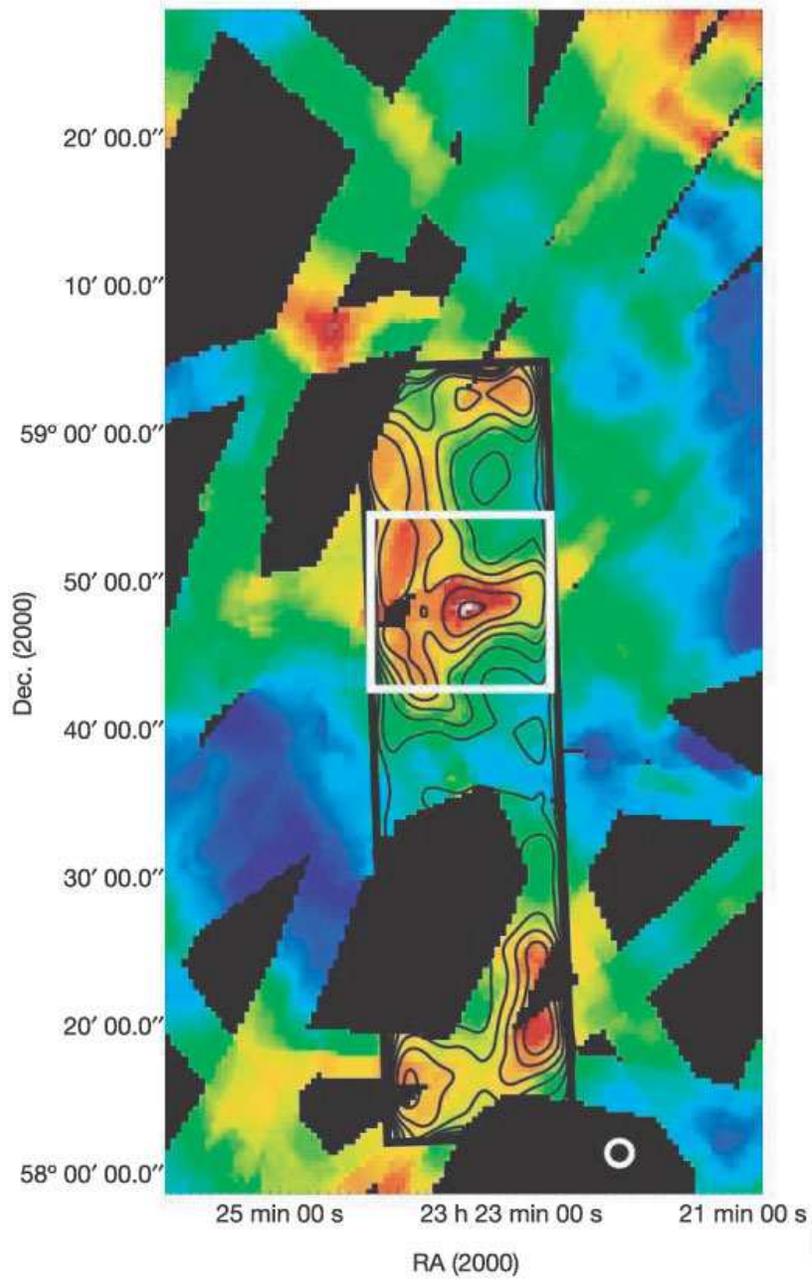

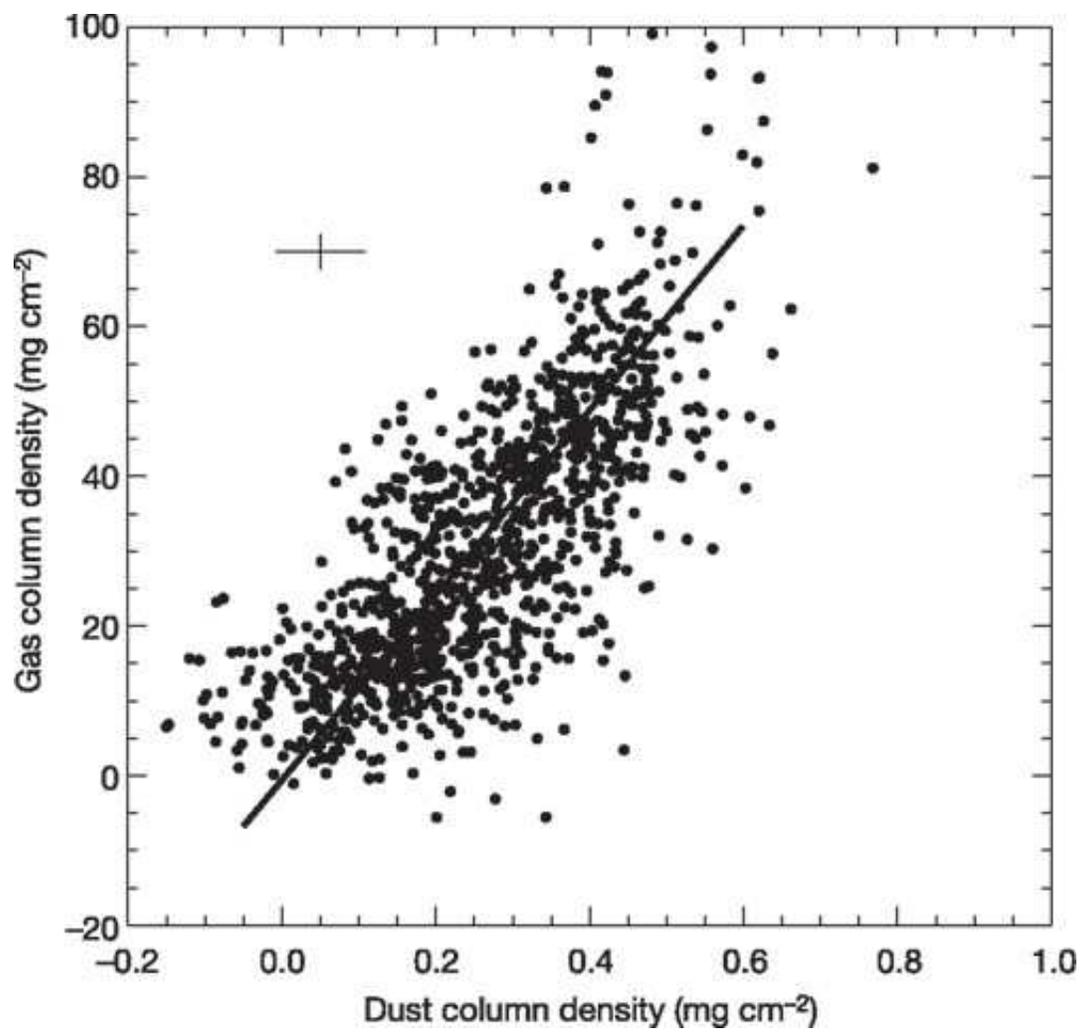